\documentclass{article}
\usepackage{graphicx}

\newtheorem{thm}{Theorem}

\begin{document}

\title{New Application of Functional Integrals to Classical Mechanics}%
\author{Anton Zherebtsov$^\dag$$^\star$, Kirill Ilinski$^\star$}%
\date{}
\maketitle
\begin{center}
{$^\dag$Division of Computational Physics, Department of Physics,
St.Petersburg State University, St.Petersburg 198904, Russia.\\
e-mail: anton.zherebtsov@fusionam.com \linebreak\linebreak
$^\star$Fusion Asset Management, 23 Berkeley Square, London W1J6 HE, UK.\\
e-mail: kirill.ilinski@fusionam.com}
\end{center}

\begin{abstract}
In this paper a new functional integral representation for
classical dynamics is introduced. It is achieved by rewriting the
Liouville picture in terms of bosonic creation-annihilation
operators and utilizing the standard derivation of functional
integrals for dynamical quantities in the coherent states
representation. This results in a new class of functional
integrals which are exactly solvable and can be found explicitly
when the underlying classical systems are integrable.
\end{abstract}

\section{Introduction}
Functional integrals are very popular in quantum mechanics and
quantum statistical physics as a basic tool of both perturbative and
non-perturbative analysis. The most fruitful applications of the
functional integrals include Faddeev-Popov ghosts in the field
theory~\cite{Faddeev}, non-perturbative corrections to the standard
perturbation theory series~\cite{instanton1,instanton2}, functional
formalism for disordered systems~\cite{efetov}, introduction of
collective variables based on Stratonovich-Hubbard
transformation(see~\cite{popov} and refs therein) and applications
to algebraic and differential
geometry~\cite{witten1,witten,polyakov}.

The beauty of the functional integrals lies in the fact that they
essentially reduce quantum mechanical problems to a classical
mechanical picture with an additional stochastic noise. This means
that instead of necessity of dealing with operators and partial
differential equations (sometimes infinitely dimensional) one can
think about the problem in terms of classical observables and, at
the final stage, to average them across possible trajectories
generated by (quantum) noise. This observation allows to find a
formal expression for any quantum observable and is a cornerstone of
all mentioned above applications.

If functional integrals bring quantum mechanics back to classics, it
is possible to formulate classical dynamics in terms of quantum one
and therefore benefit from a quantum-mechanical toolbox. The
question is not new. In last ten years series of
papers~\cite{abrikosov} examined one of possible ways to introduce
the path integral formalism into the classical mechanics. Although
bringing about interesting mathematics they, in our view, did not
result in a workable quantum-mechanical formalism which would allow
a user to employ many useful functional-integral tricks which proved
to be so helpful in the theory of many interacting degrees of
freedom, disordered systems and renormalisation group. In this paper
we are trying to fill the gap and to introduce a functional integral
representation of classical dynamics which is similar in many
respects to the standard functional integrals of many-body theory
and allows perturbative and non-perturbative treatment of classical
many-body systems and classical random dynamics. As a by-product we
also obtain class of quantum mechanical systems which are related in
our formalism to classical mechanics and therefore allow exact
solutions.

The paper is organized as follows. In the next section we briefly
remind the Liouville formulation of classical dynamics, outline the
main idea of the paper and introduce the creation-annihilation pairs
for classical momentum and coordinates. Section 2 treats in great
detail the case of one-dimensional classical harmonic oscillator. It
is shown that there exists a two-dimensional quadratic quantum
mechanical system which is equivalent to the classical oscillator in
the sense that all quantum dynamical charatecteristics can be
expressed through the corresponding classical trajectories. Section
3 builds on this and describes a general n-dimensional construction
in the case of arbitrary classical potential. This results in the
description of a new class of functional integrals and the
corresponding quantum systems which are exactly solvable and which
dynamical quantities can be found explicitly if the underlying
classical systems are integrable. Section 4 concludes the paper with
a list of possible application of the developed formalism.

\section{Classical Mechanics in the coherent representation}
Let us consider a mechanical system with the phase space
$\mathcal{M}= \{p , q\}$\footnote{To keep formulae simple in this
section we do not stress that both $p$ and $q$ can be
n-dimensional vectors.}, algebra of observables $\mathcal{A}$, set
of states $\mathcal{S}$, and Hamiltonian $H$. Assume that the
phase space $\mathcal{M}$ of the mechanical system has a volume
form invariant under the phase flow with Hamiltonian $H$. Denoting
this volume form by $dx$, we can write a probability measure
$d\mu$ as $d\mu(x) = \rho(x,t)dx$, where $\rho(x,t)$ is a positive
distribution (generalized function) on $\mathcal{M}$. For the pure
state supported at $x_0 \in \mathcal{M}$ we have $\rho(x,t) =
\delta(x - x_0)$ Dirac $\delta$-function. This representation
introduces Liouville's picture, in which observables do not depend
on time
$$\frac{df}{dt} = 0,\quad f \in \mathcal{A},$$
and states $d\mu(x) = \rho(x,t)dx$ satisfy Liouville's equation,
$$\frac{\partial\rho(x,t)}{\partial t} = -\{H, \rho\},\quad \rho(x,t)dx \in \mathcal{S},$$
which is understood in the distributional sense, and Poisson
brackets is given by:
$$
\{f,g\}=\frac{\partial f}{\partial p}\frac{\partial g}{\partial
q}-\frac{\partial f}{\partial q}\frac{\partial g}{\partial p} \ .
$$
We can rewrite this equation in Schr\"{o}dinger manner form:
\begin{equation}
\label{Liouville equation}
i\frac{\partial \rho(p,q;t)}{\partial t} = \mathcal{L}(p,q)\rho(p,q;t).
\end{equation}
where $\mathcal{L}$ is Liouville operator determined as
\begin{equation}
\label{Liouville operator}
 \mathcal{L}(p,q) = -i\frac{\partial H}{\partial p}\frac{\partial}{\partial q} + i\frac{\partial H}{\partial q}\frac{\partial}{\partial p}.
\label{liouv}
\end{equation}

Every solution of the initial value problem: $\rho(p,q;t)|_{t=0}=
\rho(p',q',0)$ for the Equation (\ref{Liouville equation}) can be
written in terms of the evolution operator $U(t) =
e^{-it\mathcal{L}}$, which is a unitary operator that relates the
state functions of the system $\rho(p,q;T)$ with $\rho(p',q';0)$.
Operator $U(t)$ is an integral operator with the kernel
$\mathcal{U}(p,q|p'q')$ and we can write:
\begin{equation}
\rho(p,q)=\int{\mathcal{U}(p,q|p',q')\rho(p',q')\,dp'dq'}.
\end{equation}

The next step is to introduce creation and annihilation operators
$a_1$, $a_2$, $a_1^\dag$, $a_2^\dag$:
$$ a_1=\frac{1}{\sqrt{2}}\left(p + \frac{\partial}{\partial p}\right), \quad a_1^\dag=\frac{1}{\sqrt{2}}\left(p - \frac{\partial}{\partial p}\right),$$
$$ a_2=\frac{1}{\sqrt{2}}\left(q + \frac{\partial}{\partial q}\right), \quad a_2^\dag=\frac{1}{\sqrt{2}}\left(q - \frac{\partial}{\partial q}\right)$$
with standard commutation relations
$$[a_1,a_1^\dag] = 1, \quad [a_2,a_2^\dag] = 1,$$
all other commutators are $0$. These operators allow us to define
the corresponding coherent states which are eigenstates of the
operators $a = (a_1$,$a_2$), $a^\dag = (a_1^\dag$,$a_2^\dag)$:
\begin{equation}
 \left|z\right> = \left|z_1z_2\right> =
 \sum_{n=0}^{\infty}\frac{\left(a_1^\dag z_1\right)^n}{n!}\sum_{m=0}^{\infty}\frac{\left(a_2^\dag z_2\right)^m}{m!}\left|0\right>\left|0\right>
 e^{-\frac{1}{2}\left| z_1\right|^2}e^{-\frac{1}{2}\left|
 z_2\right|^2} \ ,
\end{equation}
\begin{equation}
\left<\overline{z}\right| =
\left<\overline{z}_1\overline{z}_2\right| =
\left<0\right|\left<0\right| \sum_{n=0}^{\infty}\frac{\left(a_1\overline{z}_1\right)^n}{n!}\sum_{m=0}^{\infty}\frac{\left(a_2\overline{z}_2\right)^m}{m!}
 e^{-\frac{1}{2}\left|\overline{z}_1\right|^2}e^{-\frac{1}{2}\left|\overline{z}_2\right|^2}
 \ .
\end{equation}
Here $\left|0\right>$ is defined by the equation $a\left|0\right> =
0$ and unit decomposition is:
$$I=\int{\frac{dzd\overline{z}}{2\pi i}\left|z\right>\left<\overline{z}\right|} $$
It is not difficult to find the coherent states in the coordinate
representation by multiplying this equation from the left with
$\left<x\right|$ we now get $\left<x\right|a\left|0\right> = 0$, or
$$
\left<x\left|\frac{1}{\sqrt{2}}\left(x+
\frac{\partial}{\partial x}\right)\right|0\right> = 0  .
$$
This differential equation is simple to solve, and we obtain
$$
\left<x\right|\left.0\right> = \left(\frac{1}{\pi}\right)^{1/4}
e^{-x^2/2} \ .
$$
In similar manner one can find the $x$-representation of other
eigenstates $\left|n\right>$ by applying the operator $(a^\dag)^n$
to the ground state:
$$
\left<x\right|\left.n\right> =
\left<x\left|\frac{1}{\sqrt{n!}}(a^\dag)^n\right|0\right> =
\frac{H_n(x)}{\sqrt{2^n n!}\sqrt[4]{\pi}}e^{-\frac{x^2}{2}} \ .
$$
Here $H_n(x)=e^{x^2/2}\left(x-\frac{d}{dx}\right)^ne^{-x^2/2}$ is
the $n$-th Hermite polynomial. It is now straitforward to write
down the coherent states in $x$-representation:
\begin{eqnarray}
\label{<x|z>}
 \left<x\right|\left.z\right> &=&
 \left<x\left|e^{-\frac{|z|^2}{2}}\sum_{n=0}^{\infty}\frac{z^n}{\sqrt{n!}}\right|n\right>
 = e^{-\frac{|z|^2}{2}}\sum_{n=0}^{\infty}\frac{z^n H_n(x)}{n!\sqrt[4]{\pi}\sqrt{2^n}}e^{-\frac{x^2}{2}}\nonumber\\
 &=&
 \frac{1}{\sqrt[4]{\pi}}e^{-\frac{|z|^2}{2}-\frac{x^2}{2}}\sum_{n=0}^{\infty}\frac{\left(z/\sqrt{2}\right)^n}{n!}H_n(x)
 =\frac{1}{\sqrt[4]{\pi}}e^{-\frac{|z|^2}{2}-\frac{x^2}{2}}e^{-\frac{z^2}{2}+\sqrt{2}x z}
\end{eqnarray}
where we used well known formula for the generating function of Hermite polynomials:
$\sum_{n=0}^{\infty}H_n(x)\frac{s^n}{n!}=\exp\left\{-s^2 +
2sx\right\}$. The same analysis yields the following expression for the conjugated
coherent states:
\begin{equation}
\label{<z|x>}
 \left<\overline{z}\right|\left.x\right>=\frac{1}{\sqrt[4]{\pi}}\exp\left\{-\frac{|\overline{z}|^2}{2}-\frac{x^2}{2}-\frac{\overline{z}^2}{2}+\sqrt{2}x\overline{z}\right\}
\end{equation}

The coherent states are the basis for derivation of the functional
integral. As soon as the Liouville operator (\ref{liouv}) is
expresses in terms the creation-annihilation operators
$a_{1,2},a_{1,2}^{\dag}$ in the so-called Normal form (i.e. when all
creation operators stand to the left of the annihilation operators),
$\mathcal{L}_{norm}(a^\dag,a)$, one can find the evolution operator
$U(p,q|p',q')$ as
\begin{eqnarray}
 U(p,q|p',q')&=&\left<x\right|\mathcal{U}(T,0)\left|x'\right>\nonumber\\
\label{green}  &=&\int dzd\overline{z}dz'd\overline{z}'
  \left<x\right|\left.z\right> \left<\overline{z}\right|\mathcal{U}(T,0)\left|z'\right>
  \left<\overline{z}'\right|\left.x'\right>,
\end{eqnarray}
where $dzd\overline{z} = \frac{dz_1d\overline{z}_1}{2\pi i}
\frac{dz_2d\overline{z}_2}{2\pi i} $ and $dz'd\overline{z}'
=\frac{dz_1'd\overline{z}_1'}{2\pi
i}\frac{dz_2'd\overline{z}_1'}{2\pi i}$ and the kernel of evolution
operator is given by the following functional integral~\cite{popov}:
\begin{eqnarray}
 &&\left<\overline{z}_1\overline{z}_2\right|\mathcal{U}(T,0)\left|z_1'z_2'\right> = \int\mathcal{D}\psi_1 \mathcal{D}\overline{\psi}_1 \mathcal{D}\overline{\psi}_2
 \mathcal{D}\psi_2\nonumber\\
 &&\exp\left\{\sum_{k=1}^{2}\overline{\psi}_k(T)\psi_k(T)-
\frac{1}{2}|\overline{\psi}_k(T)|^2-
\frac{1}{2}|\psi_k(0)|^2\right.\nonumber\\
 &&\left.+\int\limits_0^T\left(-\overline{\psi}_1(t)\dot\psi_1(t) -
 \overline{\psi}_2(t)\dot\psi_2(t)-i
\mathcal{L}_{norm}(\overline{\psi}_1,\overline{\psi}_2,\psi_1,\psi_2)
\right)dt\right\}.
\end{eqnarray}
Here $\mathcal{D}\overline{\psi}_1\mathcal{D}\psi_1 =
\prod\limits_t\frac{d\overline{\psi}_1d\psi_1}{2\pi i}$,
$\mathcal{D}\overline{\psi}_2\mathcal{D}\psi_2 =
\prod\limits_t\frac{d\overline{\psi}_2d\psi_2}{2\pi i}$ and
initial conditions are:
\begin{eqnarray*}
\psi_1(0)=z_1', \quad \psi_2(0)=z_2'\\
\overline{\psi}_1(T)=\overline{z_1}, \quad
\overline{\psi}_2(T)=\overline{z_2}
\end{eqnarray*}
This is the functional integral representation of the classical dynamics we study
in the paper.

\section{Harmonic oscillator}
Before moving further, we would like to consider a simple example
which became rather customary to test new techniques, the harmonic
oscillator. In this section we apply the derived above functional
integral representation for the evolution operator of classical
harmonic oscillator with hamiltonian $H(p,q) =
\frac{\omega}{2}\left(p^2+q^2\right)$. The kernel of the classical
evolution operator or propagator gives us solutions of the classical
dynamical equations:
\begin{equation}
 \dot{p} = -\frac{\partial H}{\partial q},\quad \dot{q} = \frac{\partial H}{\partial p}.
\end{equation}
or
\begin{equation}
 U(q,p|q',p') = \left<q,p\;\right|\left.q',p'\right> = \delta(q-\widetilde{q}')\delta(p-\widetilde{p}'), \quad T\geq0
\end{equation}
with
\begin{eqnarray*}
 \widetilde{p}' = p'\cos\omega T - q'\sin\omega T, \quad \widetilde{q}' = q'\cos\omega T + p'\sin\omega T.
\end{eqnarray*}
and thus completely describes system behavior. We now obtain these expressions from the
functional integral for the evolution operator.

Using Equation~(\ref{Liouville operator}) one can derive the
following expression for the Liouville operator:
\begin{equation}
\label{Liouville operator for oscillatorPQ} \mathcal{L}(p,q) =
i\omega \left(q \frac{\partial}{\partial
p}-p\frac{\partial}{\partial q}\right) \ ,
\end{equation}
which can be easily expressed in the terms creation and annihilation operators
$a_1$, $a_2$, $a_1^\dag$, $a_2^\dag$:
$$ a_1=\frac{1}{\sqrt{2}}\left(p + \frac{\partial}{\partial p}\right), \quad a_1^\dag=\frac{1}{\sqrt{2}}\left(p - \frac{\partial}{\partial p}\right),$$
$$ a_2=\frac{1}{\sqrt{2}}\left(q + \frac{\partial}{\partial q}\right), \quad a_2^\dag=\frac{1}{\sqrt{2}}\left(q - \frac{\partial}{\partial q}\right)$$
as
\begin{eqnarray}
\label{Liouville operator for oscillatorAA}
\mathcal{L}(a_1,a_1^\dag,a_2,a_2^\dag)&=&\frac{i\omega}{2}\left((a_2^\dag+a_2)(a_1-a_1^\dag)-(a_1^\dag+a_1)(a_2-a_2^\dag)\right)\nonumber\\
&=&i\omega\left(a_2^\dag a_1-a_1^\dag a_2\right) =
\mathcal{L}_{Norm}(a_1,a_1^\dag,a_2,a_2^\dag).
\end{eqnarray}
This operator is already in the Normal form and can be directly used in the
functional integral for $\left<\overline{z}\left|\mathcal{U}(T,0)\right|z'\right>$:
\begin{eqnarray}
 &&\left<\overline{z}_1\overline{z}_2\right|\mathcal{U}(T,0)\left|z_1'z_2'\right> = \int\mathcal{D}\psi_1 \mathcal{D}\overline{\psi}_1 \mathcal{D}\overline{\psi}_2
 \mathcal{D}\psi_2\nonumber\\
 &&\exp\left\{\sum_{k=1}^{2}\overline{\psi}_k(T)\psi_k(T)-
\frac{1}{2}|\overline{\psi}_k(T)|^2-
\frac{1}{2}|\psi_k(0)|^2\right.\nonumber\\
 &&\left.+\int_0^T\left(-\overline{\psi}_1(t)\dot\psi_1(t) - \overline{\psi}_2(t)\dot\psi_2(t) +
 \omega \left(\overline{\psi}_2(t)\psi_1(t)-\overline{\psi}_1(t)\psi_2(t)\right)\right)dt\right\}.
\end{eqnarray}
where $\mathcal{D}\overline{\psi}_1\mathcal{D}\psi_1 = \prod\limits_t\frac{d\overline{\psi}_1d\psi_1}{2\pi i}$,
       $\mathcal{D}\overline{\psi}_2\mathcal{D}\psi_2 = \prod\limits_t\frac{d\overline{\psi}_2d\psi_2}{2\pi i}$
with the initial conditions:
\begin{eqnarray*}
 \psi_1(0)=z_1', \quad \psi_2(0)=z_2'\\
 \overline{\psi}_1(T)=\overline{z_1}, \quad \overline{\psi}_2(T)=\overline{z_2}
\end{eqnarray*}
The functional integral is Gaussian and can be easily evaluated by finding the
saddle point trajectory which are defined by the equations:
\begin{eqnarray*}
 \dot\psi_1=-\omega\psi_2, &&\quad \dot{\overline{\psi}_1}=-\omega\overline{\psi}_2,\\
 \dot\psi_2=\omega\psi_1, &&\quad \dot{\overline{\psi}_2}=\omega\overline{\psi}_1.
\end{eqnarray*}
Solutions of these equations are:
\begin{eqnarray*}
 \psi_1(t) = z_2'\cos{\omega t} - z_1'\sin{\omega t},\\
 \psi_2(t) =z_1'\cos{\omega t} + z_2'\sin{\omega t},
\end{eqnarray*}
and
\begin{eqnarray*}
 \overline{\psi}_1(t) = \overline{z}_2'\cos{\omega (t-T)} - \overline{z}_1'\sin{\omega (t-T)},\\
 \overline{\psi}_2(t) = \overline{z}_1'\cos{\omega (t-T)} + \overline{z}_2'\sin{\omega (t-T)}.
\end{eqnarray*}
Therefore the integral turns into:
\begin{eqnarray}
\label{kernel}
 \left<\overline{z}_1\overline{z}_2\right|U(T,0)\left|z_1'z_2'\right>
 = \exp\left\{(\overline{z}_1z_1'+\overline{z}_2z_2')\cos\omega T-\frac{|z_1|^2}{2}-\frac{|z_2|^2}{2}\right.\nonumber\\
 \left.-(\overline{z}_1z_2' +\overline{z}_2z_1')\sin\omega
 T-\frac{|z_1'|^2}{2}-\frac{|z_2'|^2}{2}\right\}
\end{eqnarray}
Combining together
equations~(\ref{<x|z>}),(\ref{<z|x>}),(\ref{green})
and~(\ref{kernel}) one can get the following expression for the
kernel of the evolution operator:
\begin{eqnarray*}
U(p,q|p',q')=\int\frac{dz_1d\overline{z}_1dz_2d\overline{z}_2dz_1'd\overline{z}_1'dz_2'd\overline{z}_2'}{8\pi^5}\\
\exp\left\{-|z_1|^2-|z_2|^2-|z_1'|^2-|z_2'|^2-\frac{z_1^2}{2}-\frac{z_2^2}{2}-\frac{z_2'^2}{2}-\frac{z_2'^2}{2}-\frac{p^2}{2}-\frac{q^2}{2}-\frac{p'^2}{2}-\frac{q'^2}{2}\right\}\\
\exp\left\{\sqrt{2}(z_1p+z_2q+\overline{z}'_1p'+\overline{z}'_2q') +
(\overline{z}_1z'+\overline{z}_2z')\cos\omega T -
(\overline{z}_1'z_2'+\overline{z}_1'z_2')\sin\omega T\right\}.
\end{eqnarray*}
Now it is possible to show that this is Dirac
$\delta$-function on the solutions of the classical mechanical equations of motion.
Indeed, consider the following decomposition:

\begin{eqnarray*}
 \delta(x-x')=\left<x\right|I\left|x'\right> = \int{\left<x\right|\left.z\right>\left<\overline{z}\right|\left.z'\right>\left<\overline{z'}\right|\left.x'\right>\frac{dzd\overline{z}dz'd\overline{z'}}{4i\pi i\pi}}\\
 =\int{\frac{dzd\overline{z}dz'd\overline{z'}}{4i\pi i\pi}\frac{1}{\sqrt{\pi}}\exp\left\{-\frac{|z|^2}{2}-\frac{x^2}{2}-\frac{z^2}{2}+\sqrt{2}x z\right\}}\times\\
 \exp\left\{-\frac{|\overline{z'}|^2}{2}-\frac{x'^2}{2}-\frac{\overline{z'}^2}{2}+\sqrt{2}x'\overline{z'}-\frac{|z|^2}{2}-\frac{|z'|^2}{2}-\overline{z}z'\right\}.
\end{eqnarray*}
Using this decomposition and introducing a set of new variables:
\begin{eqnarray*}
 \widetilde{z}'_1 = z'_1\cos\omega T - z'_2\sin\omega T, \quad \overline{\widetilde{z}}'_1 = \overline{z}'_1\cos\omega T - \overline{z}'_2\sin\omega T, \quad\\
 \widetilde{z}'_2 = z'_2\cos\omega T + z'_1\sin\omega T, \quad \overline{\widetilde{z}}'_1 = \overline{z}'_1\cos\omega T + \overline{z}'_2\sin\omega T, \quad
\end{eqnarray*}
and
\begin{eqnarray*}
 \widetilde{p}' = p'\cos\omega T - q'\sin\omega T, \quad \widetilde{q}' = q'\cos\omega T + p'\sin\omega T.
\end{eqnarray*}
the kernel transforms into
\begin{eqnarray*}
 U(p,q|p',q')=\int\frac{dz_1d\overline{z}_1dz_2d\overline{z}_2d\widetilde{z}_1'd\overline{\widetilde{z}}_1'd\widetilde{z}_2'd\overline{\widetilde{z}}_2'}{\pi^5}\times\\
 \exp\left\{-|z_1|^2-|z_2|^2-|\widetilde{z_1}|^2-|\widetilde{z_2}'|^2\right\}\times\\
 \exp\left\{-\frac{z_1^2}{2}-\frac{z_2^2}{2}-\frac{z_2'^2}{2}-\frac{z_2'^2}{2}-\frac{p^2}{2}-\frac{q^2}{2}-\frac{\widetilde{p}'^2}{2}-\frac{\widetilde{q}'^2}{2}\right\}\times\\
 \exp\left\{\sqrt{2}(z_1p+z_2q+\overline{\widetilde{z}}'_1\widetilde{p}'+\overline{\widetilde{z}}'_2\widetilde{q}') + (\overline{z}_1\widetilde{z}'_1+\overline{z}_2\widetilde{z}'_2)\right\}.
\end{eqnarray*}
which produces the required result
\begin{eqnarray*}
 U(p,q|p',q')=\delta(q-\widetilde{q}')\delta(p-\widetilde{p}') \ .
\end{eqnarray*}
This shows that the functional integral representation for the
matrix elements of classical evolution operator indeed reproduces
the Liouville flow on the classical phase space and is equivalent
to the solution of classical Hamilton equations.

\section{New class of localization formulas for functional integrals}
In this section we describe a new class of functional integrals
which can be evaluated exactly using their relationship to some
classical dynamical systems. The main result is summarized in the
following
\begin{thm}
For any function $V(q_1\ldots q_n)$ the following integral:
\begin{eqnarray}
I_V(z,\overline{z})=\int\mathcal{D}\psi_1
\mathcal{D}\overline{\psi}_1 \mathcal{D}\overline{\psi}_2
\mathcal{D}\psi_2
\exp\left\{\sum_{i=1}^{n}\sum_{k=1}^{2}\overline{\psi}_k^i(T)\psi_k^i(T)-
\frac{1}{2}|\overline{\psi}_k^i(T)|^2-\frac{1}{2}|\psi_k^i(0)|^2 \right\}\nonumber\\
\exp\left\{\int_0^T{\sum_{i=1}^n
-\overline{\psi}_1^i(t)\dot\psi_1^i(t) -
\overline{\psi}_2^i(t)\dot\psi_2^i(t) +
\left(\overline{\psi}_1^i(t)\psi_2^i(t) -
\overline{\psi}_2^i(t)\psi_1^i(t) \right)}\right. \nonumber\\+
\left.(\psi_1^{i}(t)-\overline{\psi}^{i}_1(t))\mathcal{L}_{i,norm}^{(V)}(\psi,\overline{\psi})\right\}
\end{eqnarray}
where $\mathcal{D}\overline{\psi}_1\mathcal{D}\psi_1 =
\prod\limits_t\frac{d\overline{\psi}_1d\psi_1}{2\pi i}$,
$\mathcal{D}\overline{\psi}_2\mathcal{D}\psi_2 =
\prod\limits_t\frac{d\overline{\psi}_2d\psi_2}{2\pi i}$ with the
initial conditions:
\begin{eqnarray*}
\psi_1(0)=z_1', \quad \psi_2(0)=z_2'\\
\overline{\psi}_1(T)=\overline{z_1}, \quad
\overline{\psi}_2(T)=\overline{z_2}
\end{eqnarray*}
and $ \mathcal{L}_{i,norm}(a,a^{\dag}) = N\left(
\left.\frac{\partial V}{\partial q^i}
-q^i\right)\right|_{q^i=\frac{1}{\sqrt{2}}(a_2^i+a_2^{\dag i})} \ ,
$ \\is equal to
\begin{eqnarray}
I_V(z,\overline{z})=\int\prod_{i=1}^{n}dp'^idq'^i
 \left(\frac{1}{\sqrt{\pi}}e^{-\frac{p^i(T)^2}{2}-\frac{q^i(T)^2}{2}-\frac{p'^{i2}}{2}-\frac{q'^{i2}}{2}}
 e^{-\frac{|z_1'^i|^2}{2}-\frac{|z_2'^i|^2}{2}-\frac{|\overline{z}_1^i|^2}{2}-\frac{|\overline{z}_2^i|^2}{2}}\right.\nonumber\\
 \left. e^{-\frac{z_1'^{i2}}{2}-\frac{z_2'^{i2}}{2}-\frac{\overline{z}_1^{i2}}{2}-\frac{\overline{z}_2^{i2}}{2}}
 e^{\sqrt{2}(z_1'^ip'^i+z_2'^iq'^i+\overline{z}_1^ip^i(T)+\overline{z}_2q^i(T))}\right) \
 ,
\end{eqnarray}
where $\{p^{i}(T),q^{i}(T)\}$ are the solution of the classical
Hamiltonian equations with the Hamilton function
$$
H=\frac{1}{2}\sum_{i=1}^{n} p^{i2} + V(q^1\ldots q^n)\ .
$$
for the time $t=T$ with initial conditions at $t=0$ being
$\{p'^{i},q'^{i}\}$.
\end{thm}
\textit{Proof.} Let us consider classical system with the
following Hamilton function:
$$
H=\frac{1}{2}\sum_{i=1}^{n} p^{i2} + V(q^1\ldots q^n) =
\frac{1}{2}\sum_{i=1}^{n}\left(p^{i2} + q^{i2}\right)+ V(q^1\ldots
q^n)-\frac{1}{2}\sum_{i=1}^{n}q^{i2} \ .
$$
The corresponding Liouville operator $\mathcal{L}$ will have the
form:
\begin{equation}
\mathcal{L} = i\sum_{i=1}^{n}\left(q^i\frac{\partial}{\partial
 p^i}-p^i\frac{\partial}{\partial
q^i}\right) + i\sum_{i=1}^{n}\left(\frac{\partial V}{\partial
q^i}-q^i\right)\frac{\partial}{\partial p^i} \ .
\end{equation}
Expressing this operator in terms of creation-annihilation
operators $a_1^{i}, a_2^{i}, a_1^{\dag i}, a_2^{\dag i}$
\begin{eqnarray}
 a_1^i=\frac{1}{\sqrt{2}}\left(p^i + \frac{\partial}{\partial p^i}\right), \quad a_1^{i\dag}=\frac{1}{\sqrt{2}}\left(p^i-\frac{\partial}{\partial p^i}\right),\\
 a_2^i=\frac{1}{\sqrt{2}}\left(q^i + \frac{\partial}{\partial q^i}\right), \quad a_2^{i\dag}=\frac{1}{\sqrt{2}}\left(q^i-\frac{\partial}{\partial q^i}\right),
\end{eqnarray}
one can get
\begin{eqnarray}
\mathcal{L} = i\sum_{i=1}^{n}\left(a_2^{\dag i}a_1^{i}-a_1^{\dag
i}a_2^{i}\right) + \frac{i}{\sqrt{2}}\sum_{i=1}^{n} a_1^i\left(
\left.\frac{\partial V}{\partial q^i}
-q^i\right)\right|_{q^i=\frac{1}{\sqrt{2}}(a_2^i+a_2^{\dag
i})}\nonumber\\
 \label{Liouv}-\frac{i}{\sqrt{2}}\sum_{i=1}^{n} a_{1}^{\dag i} \left(
\left.\frac{\partial V}{\partial
q^i}-q^i\right)\right|_{q^i=\frac{1}{\sqrt{2}}(a_2^i+a_2^{\dag i})}
\ .
\end{eqnarray}
If the this equation is rewritten in the Normal form:
\begin{eqnarray}
\mathcal{L} = i\sum_{i=1}^{n}\left(a_2^{\dag i}a_1^{i}-a_1^{\dag
i}a_2^{i}\right) + \frac{i}{\sqrt{2}}\sum_{i=1}^{n}
\mathcal{L}_{i,norm}(a,a^{\dag})a_{1}^{i}\nonumber\\
\label{Liouv1}-\frac{i}{\sqrt{2}}\sum_{i=1}^{n} a_{1}^{\dag i}
\mathcal{L}_{i,norm}(a,a^{\dag}) \ ,
\end{eqnarray}
where
$$
\mathcal{L}_{i,norm}(a,a^{\dag}) = N\left( \left.\frac{\partial
V}{\partial q^i}
-q^i\right)\right|_{q^i=\frac{1}{\sqrt{2}}(a_2^i+a_2^{\dag i})} \,
$$
it is straightforward to put down the functional integral for the
matrix elements of the evolution operator in the coherent state
representation:
\begin{eqnarray}
\left<\overline{z}_1\overline{z}_2\right|\mathcal{U}(T,0)\left|z_1'z_2'\right>
= \int\mathcal{D}\psi_1 \mathcal{D}\overline{\psi}_1
\mathcal{D}\overline{\psi}_2 \mathcal{D}\psi_2 \nonumber\\
\exp\left\{\sum_{i=1}^{n}\sum_{k=1}^{2}\overline{\psi}_k^i(T)\psi_k^i(T)-
\frac{1}{2}|\overline{\psi}_k^i(T)|^2-\frac{1}{2}|\psi_k^i(0)|^2 \right\}\nonumber\\
\exp\left\{\int_0^T{\sum_{i=1}^n
-\overline{\psi}_1^i(t)\dot\psi_1^i(t) -
\overline{\psi}_2^i(t)\dot\psi_2^i(t) + \left(
\overline{\psi}_2^i(t)\psi_1^i(t)
-\overline{\psi}_1^i(t)\psi_2^i(t)\right)} \right.\nonumber\\-
\left.(\psi_1^{i}(t)-\overline{\psi}^{i}_1(t))\mathcal{L}_{i,norm}(\psi,\overline{\psi})dt\right\}
\end{eqnarray}
where $\mathcal{D}\overline{\psi}_1\mathcal{D}\psi_1 =
\prod\limits_t\prod\limits_{l=1}^n\frac{d\overline{\psi}_1^ld\psi_1^l}{2\pi
i}$, $\mathcal{D}\overline{\psi}_2\mathcal{D}\psi_2 =
\prod\limits_t\prod\limits_{l=1}^n\frac{d\overline{\psi}_2^ld\psi_2^l}{2\pi
i}$ with the initial conditions:
\begin{eqnarray*}
\psi_1(0)=z_1', \quad \psi_2(0)=z_2'\\
\overline{\psi}_1(T)=\overline{z_1}, \quad
\overline{\psi}_2(T)=\overline{z_2}
\end{eqnarray*}
The same matrix element can be found as
\begin{eqnarray}
 U(\{\overline{z}_1^i,\overline{z}_2^i\}|\{z_1'^i,z_2'^i\})=\left<\{\overline{z}_1^i,\overline{z}_2^i\}\left|\mathcal{U}(0,T)\right|\{z_1'^i,z_2'^i\}\right>\nonumber\\
 =\int\prod_{i=1}^{n}dp^idp'^idq^idq'^i\left<\{\overline{z}_1^i,\overline{z}_2^i\}\left|\right.\{p^i,q^i\}\right>\left<\{p'^i,q'^i\}\left|\right.\{z_1'^i,z_2'^i\}\right>\times\nonumber\\
 \left<\{p^i,q^i\}\left|\mathcal{U}(0,T)\right|\{p'^i,q'^i\}\right> \ .
\end{eqnarray}
Using equations~(\ref{<x|z>}) and~(\ref{<z|x>}) one can easily
get:
\begin{eqnarray}
\label{MainInt}
 U(\{\overline{z}_1^i,\overline{z}_2^i\}|\{z_1'^i,z_2'^i\})\nonumber\\
 =\int\prod_{i=1}^{n}dp^idp'^idq^idq'^i\left<\{p^i,q^i\}\left|\mathcal{U}(0,T)\right|\{p'^i,q'^i\}\right>\times\nonumber\\
 \mathcal{R}\left(\{p^i,q^i\},\{p'^i,q'^i\},\{z_1'^i,z_2'^i\},\{\overline{z}_1^i,\overline{z}_2^i\}\right),
\end{eqnarray}
where
\begin{eqnarray}
 &&\mathcal{R}\left(\{p^i,q^i\},\{p'^i,q'^i\},\{z_1'^i,z_2'^i\},\{\overline{z}_1^i,\overline{z}_2^i\}\right)\nonumber\\
 &&=\left<\{\overline{z}_1^i,\overline{z}_2^i\}\left|\right.\{p^i,q^i\}\right>\left<\{p'^i,q'^i\}\left|\right.\{z_1'^i,z_2'^i\}\right>\nonumber\\
 &&=\prod_{i=1}^{n}\left(\frac{1}{\sqrt{\pi}}e^{-\frac{p^{i2}}{2}-\frac{q^{i2}}{2}-\frac{p'^{i2}}{2}-\frac{q'^{i2}}{2}}
 e^{-\frac{|z_1'^i|^2}{2}-\frac{|z_2'^i|^2}{2}-\frac{|\overline{z}_1^i|^2}{2}-\frac{|\overline{z}_2^i|^2}{2}}\times\right.\nonumber\\
 &&\left.e^{-\frac{z_1'^{i2}}{2}-\frac{z_2'^{i2}}{2}-\frac{\overline{z}_1^{i2}}{2}-\frac{\overline{z}_2^{i2}}{2}}
 e^{\sqrt{2}(z_1'^ip'^i+z_2'^iq'^i+\overline{z}_1^ip^i+\overline{z}_2^iq^i)}\right) \ .
\end{eqnarray}
Taking into account that the classical evolution operator is a
delta-function on the solution of the classical equations it is easy
to see that expression~(\ref{MainInt}) can be re-arranged as:
\begin{eqnarray*}
 U(\{\overline{z}_1^i,\overline{z}_2^i\}|\{z_1'^i,z_2'^i\})
 =\int\prod_{i=1}^{n}dp'^idq'^i
 \left(\frac{1}{\sqrt{\pi}}e^{-\frac{p^i(T)^2}{2}-\frac{q^i(T)^2}{2}-\frac{p'^{i2}}{2}-\frac{q'^{i2}}{2}}\right.\nonumber\\
 \left.e^{-\frac{|z_1'^i|^2}{2}-\frac{|z_2'^i|^2}{2}-\frac{|\overline{z}_1^i|^2}{2}-\frac{|\overline{z}_2^i|^2}{2}}
 e^{-\frac{z_1'^{i2}}{2}-\frac{z_2'^{i2}}{2}-\frac{\overline{z}_1^{i2}}{2}-\frac{\overline{z}_2^{i2}}{2}}
 e^{\sqrt{2}(z_1'^ip'^i+z_2'^iq'^i+\overline{z}_1^ip^i(T)+\overline{z}_2q^i(T))}\right) \ .
\end{eqnarray*}
where $\{p^{i}(T),q^{i}(T)\}$ are the solution of the classical
Hamiltonian equations for the time $t=T$ with initial conditions
at $t=0$ being $\{p'^{i},q'^{i}\}$. This gives the formula for
exact evaluation of the described class of functional integrals in
terms of solutions of the corresponding classical equations of
motion. In the case of integrable systems when the solutions can
be written down explicitly, the functional integrals will be
evaluated explicitly as well. This completes the proof.

\section{Concluding Remarks}
In this paper we suggested a new functional integral representation for classical
evolution operator. Having calculated this expression one can study time evolution of
any classical observable as well as statistical properties of classical systems.
As a by-product we also found a set of quantum mechanical systems which are
related to the classical mechanics and can be solved exactly by reducing the problem to
solutions of the corresponding classical equations.

As we pointed in the Introduction the existence of the functional
integral representation opens interesting technical opportunities.
Let us list here the main directions which, we believe, promise
interesting results in future:
\begin{enumerate}
\item Study of classical dynamical systems in the presence of
random forces. Since the evolution operator can be formally
written for any interaction structure, one can derive directly a
functional integral representation for averaged dynamics of
classical observables as well as their statistical characteristics
in close analogy with the corresponding quantum
case~\cite{efetov}. One of applications in this case would be
study of classical chaotic systems by means of non-perturbative
field-theoretical methods.

\item Collective variables in classical dynamical systems. One can
apply functional form of Stratonovich-Hubbard transformation to
introduce collective variables into the classical systems with
many interacting degrees of freedom. This might be an interesting
bridge between classical n-particle dynamics and classical
statistical physics.

\item Investigation of possibility to introduce dissipative forces
into the formalism. This can potentially lead to a new instrument
of study non-equilibrium open systems and self-organization.

\item Exact solutions of quantum mechanical or field theoretical
problems in terms of the related classical trajectories. One can
see the relation established here between as certain class of
quantum systems and some classical systems as a form of
localization formulas for the corresponding quantum functional
integrals. It is interesting to see how wide is the class and if
the formalism can be generalized to include even more quantum
systems (not necessary 2n-dimensional ones).
\end{enumerate}
We hope to return to these issues in our future publications.\\

\noindent {\bf Acknowledgement} We would like to thank good old
Sasha Stepanenko for discussion of the subject in 1992. We also
grateful to Serge Levin for questions and comments.


\begin{thebibliography}{99}
\bibitem{Faddeev} L.D.Faddeev, V.N.Popov, Phys. Lett. B {\bf 25} (1978)
\bibitem{instanton1} 't.Hooft G., Phys. Rev. Lett. {\bf 37}:8 (1976)
\bibitem{instanton2} R.Rajaraman, "Solitons and Instantons", Noth-Holland Publishing Co. (1982)
\bibitem{efetov} K.B.Efetov, "Supersymmetry in Disorder and Chaos", Cambridge University Press, New York. (1997)
\bibitem{popov} V.N.Popov, "Functional integrals and collective exitations", Cambrige University Press. (1987)
\bibitem{witten1} E. Witten, Quantum Field Theory And The Jones Polynomial, Comm. Math. Phys. 121 (1989) 351.
\bibitem{witten} Avarez-Gaume, L: Supersymmetry and the Atiyah-Singer Index Theorem,Commun.Math.Phys {\bf 90}, 161-173 (1983)
\bibitem{polyakov} N.V.Borisov, K.N.Ilinski, N=2 Supersymmetric Quantum
Mechanics onRiemann surfaces with meromorphic superpotentials,
Commun.Math.Phys {\bf 161}, 177-194 (1994); N.V.Borisov,
K.N.Ilinski, G.V.Kalinin, New index formulas as a meromorphic
generalization of Chern-Gauss-Bonnet theorem,Lett.Math.Phys. {\bf
20},  21-27, (1997)
\bibitem{abrikosov} E.Gozzi, M.Reuter,W.D.Thacker, Phys. Rev.D {\bf 40} (1989) 3363;
A.A.Abrikosov (Jr.), Nucl. Phys. B {\bf 382} (1992) 581; E.Deotto
and E.Gozzi and D.Mauro, Hilbert Space Structure in Classical
Mechanics: (I), Rend. Sem. Mat. Univ. Politec. Torino {\bf 54},
269-277 (1996); E.Gozzi and M.Regini, Addenda and corrections to
work done on the path-integral approach to classical mechanics,
Phys. Rev. D {\bf 62} 067702 (2000); Hilbert Space Structure in
Classical Mechanics: (II), J. Math. Phys. {\bf 44} 5937-5957
(2003);E.Gozzi, Functional Techniques in Classical Mechanics, Nucl.
Phys. Proc. Suppl. {\bf 104} (2001); A.A.Abrikosov(jr) and E.Gozzi
and D.Mauro, Time and Geometric Quantization, Mod. Phys. Lett. A18,
2347-2354 (2003)
\end{thebibliography}

\end{document}